\documentclass[aps,prl,twocolumn,tightenlines,superscriptaddress,showpacs,byrevtex]{revtex4}

\usepackage{graphicx}
\usepackage{epsfig}
\usepackage{dcolumn}
\usepackage{bm}
\usepackage{rotating}

\newcommand{\lum}{{\cal L}}

\newcommand{\BR}{{\cal B}}

\newcommand{\piz}{\pi^0}

\newcommand{\ks}{K_S^0}

\newcommand{\psp}{\psi(2S)}
\newcommand{\jpsi}{J/\psi}
\newcommand{\EE}{e^+e^-}
\newcommand{\MM}{\mu^+\mu^-}

\newcommand{\pp}{\pi^+\pi^-}

\newcommand{\kk}{K^+K^-}

\newcommand{\beq}{\begin{equation}}
\newcommand{\eeq}{\end{equation}}
\newcommand{\beqy}{\begin{eqnarray}}
\newcommand{\eeqy}{\end{eqnarray}}
\newcommand{\bitm}{\begin{itemize}}
\newcommand{\eitm}{\end{itemize}}

\def\Journal#1#2#3#4{{#1} {\bf #2}, #3 (#4)}

\def\PRD{Phys. Rev. D}

\begin{document}

\preprint{} \preprint{\vbox{
        \hbox{Belle Preprint 2013-8}
        \hbox{KEK   Preprint 2013-6}
        \hbox{Intended for {\it Phys. Rev. D(R)}}
        \hbox{Authors: C. P. Shen, C. Z. Yuan }
        \hbox{Committee: S. Eidelman(chair), R. Leitner, G. Tatishvili}
         }}

\title{\quad \\[-0.1cm]
Measurement of exclusive $\Upsilon(1S)$ and $\Upsilon(2S)$ decays
into Vector-Pseudoscalar final states}



\noaffiliation
\affiliation{University of the Basque Country UPV/EHU, 48080 Bilbao}
\affiliation{Beihang University, Beijing 100191}
\affiliation{University of Bonn, 53115 Bonn}
\affiliation{Budker Institute of Nuclear Physics SB RAS and Novosibirsk State University, Novosibirsk 630090}
\affiliation{Faculty of Mathematics and Physics, Charles University, 121 16 Prague}
\affiliation{University of Cincinnati, Cincinnati, Ohio 45221}
\affiliation{Deutsches Elektronen--Synchrotron, 22607 Hamburg}
\affiliation{Justus-Liebig-Universit\"at Gie\ss{}en, 35392 Gie\ss{}en}
\affiliation{Gifu University, Gifu 501-1193}
\affiliation{II. Physikalisches Institut, Georg-August-Universit\"at G\"ottingen, 37073 G\"ottingen}
\affiliation{Hanyang University, Seoul 133-791}
\affiliation{University of Hawaii, Honolulu, Hawaii 96822}
\affiliation{High Energy Accelerator Research Organization (KEK), Tsukuba 305-0801}
\affiliation{Hiroshima Institute of Technology, Hiroshima 731-5193}
\affiliation{Ikerbasque, 48011 Bilbao}
\affiliation{Indian Institute of Technology Guwahati, Assam 781039}
\affiliation{Indian Institute of Technology Madras, Chennai 600036}
\affiliation{Institute of High Energy Physics, Chinese Academy of Sciences, Beijing 100049}
\affiliation{Institute of High Energy Physics, Vienna 1050}
\affiliation{Institute for High Energy Physics, Protvino 142281}
\affiliation{INFN - Sezione di Torino, 10125 Torino}
\affiliation{Institute for Theoretical and Experimental Physics, Moscow 117218}
\affiliation{J. Stefan Institute, 1000 Ljubljana}
\affiliation{Kanagawa University, Yokohama 221-8686}
\affiliation{Institut f\"ur Experimentelle Kernphysik, Karlsruher Institut f\"ur Technologie, 76131 Karlsruhe}
\affiliation{Korea Institute of Science and Technology Information, Daejeon 305-806}
\affiliation{Korea University, Seoul 136-713}
\affiliation{Kyungpook National University, Daegu 702-701}
\affiliation{\'Ecole Polytechnique F\'ed\'erale de Lausanne (EPFL), Lausanne 1015}
\affiliation{Faculty of Mathematics and Physics, University of Ljubljana, 1000 Ljubljana}
\affiliation{University of Maribor, 2000 Maribor}
\affiliation{Max-Planck-Institut f\"ur Physik, 80805 M\"unchen}
\affiliation{School of Physics, University of Melbourne, Victoria 3010}
\affiliation{Moscow Physical Engineering Institute, Moscow 115409}
\affiliation{Graduate School of Science, Nagoya University, Nagoya 464-8602}
\affiliation{Kobayashi-Maskawa Institute, Nagoya University, Nagoya 464-8602}
\affiliation{Nara Women's University, Nara 630-8506}
\affiliation{National Central University, Chung-li 32054}
\affiliation{National United University, Miao Li 36003}
\affiliation{Department of Physics, National Taiwan University, Taipei 10617}
\affiliation{H. Niewodniczanski Institute of Nuclear Physics, Krakow 31-342}
\affiliation{Nippon Dental University, Niigata 951-8580}
\affiliation{Niigata University, Niigata 950-2181}
\affiliation{University of Nova Gorica, 5000 Nova Gorica}
\affiliation{Osaka City University, Osaka 558-8585}
\affiliation{Pacific Northwest National Laboratory, Richland, Washington 99352}
\affiliation{Panjab University, Chandigarh 160014}
\affiliation{University of Pittsburgh, Pittsburgh, Pennsylvania 15260}
\affiliation{Research Center for Electron Photon Science, Tohoku University, Sendai 980-8578}
\affiliation{University of Science and Technology of China, Hefei 230026}
\affiliation{Seoul National University, Seoul 151-742}
\affiliation{Soongsil University, Seoul 156-743}
\affiliation{Sungkyunkwan University, Suwon 440-746}
\affiliation{School of Physics, University of Sydney, NSW 2006}
\affiliation{Tata Institute of Fundamental Research, Mumbai 400005}
\affiliation{Excellence Cluster Universe, Technische Universit\"at M\"unchen, 85748 Garching}
\affiliation{Toho University, Funabashi 274-8510}
\affiliation{Tohoku Gakuin University, Tagajo 985-8537}
\affiliation{Tohoku University, Sendai 980-8578}
\affiliation{Department of Physics, University of Tokyo, Tokyo 113-0033}
\affiliation{Tokyo Institute of Technology, Tokyo 152-8550}
\affiliation{Tokyo Metropolitan University, Tokyo 192-0397}
\affiliation{Tokyo University of Agriculture and Technology, Tokyo 184-8588}
\affiliation{University of Torino, 10124 Torino}
\affiliation{CNP, Virginia Polytechnic Institute and State University, Blacksburg, Virginia 24061}
\affiliation{Wayne State University, Detroit, Michigan 48202}
\affiliation{Yamagata University, Yamagata 990-8560}
\affiliation{Yonsei University, Seoul 120-749}
 \author{C.~P.~Shen}\affiliation{Beihang University, Beijing 100191}\affiliation{Graduate School of Science, Nagoya University, Nagoya 464-8602}
  \author{C.~Z.~Yuan}\affiliation{Institute of High Energy Physics, Chinese Academy of Sciences, Beijing 100049} 
  \author{I.~Adachi}\affiliation{High Energy Accelerator Research Organization (KEK), Tsukuba 305-0801} 
  \author{H.~Aihara}\affiliation{Department of Physics, University of Tokyo, Tokyo 113-0033} 
  \author{D.~M.~Asner}\affiliation{Pacific Northwest National Laboratory, Richland, Washington 99352} 
  \author{V.~Aulchenko}\affiliation{Budker Institute of Nuclear Physics SB RAS and Novosibirsk State University, Novosibirsk 630090} 
  \author{A.~M.~Bakich}\affiliation{School of Physics, University of Sydney, NSW 2006} 
  \author{A.~Bala}\affiliation{Panjab University, Chandigarh 160014} 
  \author{B.~Bhuyan}\affiliation{Indian Institute of Technology Guwahati, Assam 781039} 
  \author{M.~Bischofberger}\affiliation{Nara Women's University, Nara 630-8506} 
% ...
\collaboration{The Belle Collaboration}

\begin{abstract}

Using samples of 102 million $\Upsilon(1S)$ and 158 million
$\Upsilon(2S)$ events collected with the Belle detector, we study
exclusive hadronic decays of these two bottomonium resonances to
$\ks K^+ \pi^-$ and charge-conjugate (c.c.) states, $\pi^+ \pi^- \pi^0
\pi^0$, and $\pi^+ \pi^- \pi^0$, and to the two-body
Vector-Pseudoscalar ($K^{\ast}(892)^0\bar{K}^0+ {\rm c.c.}$,
$K^{\ast}(892)^-K^++ {\rm c.c.}$, $\omega\pi^0$, and $\rho\pi$) final states.
For the first time, signals are observed
in the modes $\Upsilon(1S)\to \ks K^+
\pi^- + {\rm c.c.}$, $\pi^+ \pi^- \pi^0 \pi^0$, and $\Upsilon(2S) \to \pi^+ \pi^- \pi^0
\pi^0$, and evidence is found for the modes $\Upsilon(1S)\to \pi^+ \pi^- \pi^0$,
$K^{\ast }(892)^0 \bar{K}^0+ {\rm c.c.}$, and  $\Upsilon(2S) \to \ks K^+
\pi^- + {\rm c.c.}$  Branching fractions are measured for all the
processes, while 90\% confidence level
upper limits on the branching fractions are also set for
the modes with a
statistical significance of less than $3\sigma$. The ratios of the
branching fractions of $\Upsilon(2S)$ and $\Upsilon(1S)$ decays
into the same final state are used to test a perturbative QCD
prediction for OZI-suppressed bottomonium decays.

\end{abstract}


\maketitle


The $\Upsilon(1S)$ and $\Upsilon(2S)$ are expected to decay mainly
via three gluons, with a few percent probability
 to two gluons and a photon~\cite{PDG}.
The two- and three-gluon channels provide an entry to many potential
final states, including states made of pure glue (glueballs), light
Higgs bosons, and states made of light quarks. The study of $\Upsilon(1S)$ and $\Upsilon(2S)$
hadronic decays may pave the way for a more complete understanding
of how gluon final states fragment into hadrons.
However, little experimental information is available on exclusive
decays of the $\Upsilon$ resonances below $B\bar{B}$ threshold.
Recently, a few Vector-Tensor (VT) and Axial-vector-Pseudoscalar
states from $\Upsilon(1S)$ and $\Upsilon(2S)$
decays were measured by the Belle Collaboration~\cite{shen}.

Perturbative quantum chromodynamics (pQCD) provides a relation for
the ratios of the branching fractions ($\cal B$) for the OZI
(Okubo-Zweig-Iizuka)~\cite{OZI} suppressed $\jpsi$ and $\psi(2S)$
decays to hadrons~\cite{appelquist}
\begin{equation}
Q_{\psi}=\frac{{\cal B}_{\psi(2S) \to { {\rm hadrons}}}}{{\cal
B}_{J/\psi \to {\rm hadrons}}} =\frac{{\cal B}_{\psi(2S) \to
e^+e^-}}{{\cal B}_{J/\psi \to e^+e^-}} \approx 12\%,
\end{equation}
which is referred to as the ``12\% rule'' and is expected to apply
with reasonable accuracy to both inclusive and exclusive decays.
However, substantial deviations are seen for $\rho\pi$ and
other Vector-Pseudoscalar (VP) final states such as
$K^{\ast}(892)\bar{K}$, as well as for
VT final states~\cite{nakao}. This is the so-called
``$\rho\pi$ puzzle.'' None of the many existing theoretical
explanations that have been proposed have been able to accommodate
all of the measurements reported to date~\cite{myw_review}.

A similar rule can be derived for OZI-suppressed bottomonium
decays, where we expect
 \begin{equation}
Q_{\Upsilon} =\frac{{\cal B}_{\Upsilon(2S) \to {\rm hadrons}}}{{\cal
B}_{\Upsilon(1S) \to {\rm hadrons}}} = \frac{{\cal B}_{\Upsilon(2S) \to
e^+e^-}}{{\cal B}_{\Upsilon(1S) \to e^+e^-}} = 0.77\pm 0.07.
 \end{equation}
This rule should hold better than the 12\% rule for charmonium
decay since the bottomonium states have higher mass and pQCD and the
potential models have better predictive power, as has been
demonstrated in calculations of the $b\bar{b}$ meson spectrum.
For the $\pp \pi^0$ and $\rho \pi$  modes, upper limits of
$1.84 \times 10^{-5}$ and $2\times 10^{-4}$ have been published~\cite{fulton}
for the decays $\Upsilon(1S) \to \pp \pi^0$ and  $\Upsilon(1S) \to \rho \pi$,
respectively.

If violation of the pQCD rules is observed in the bottomonium system, a comparison
with the charmonium system may help to
develop a theoretical explanation of the $\rho\pi$ puzzle.
For $K^{\ast}(892) {\bar K}$, there is a large
isospin-violating difference between the branching fractions for
the charged and neutral $\psp \to K^{\ast}(892) {\bar K}$ decays;
this is not seen in $\jpsi$ decays~\cite{PDG}.
This pattern can be probed in $\Upsilon(1S)$ and
$\Upsilon(2S)$ decays.


In this paper, we report studies of exclusive hadronic decays of
the $\Upsilon(1S)$ and $\Upsilon(2S)$ resonances to the
$\ks K^{+} \pi^-$~\cite{charge}, $\pp \pi^0 \pi^0$, and
$\pp \pi^0$, and two-body VP ($K^{\ast}(892)^0\bar{K}^0$,
$K^{\ast}(892)^-K^+$, $\omega\pi^0$, and $\rho\pi$) final states.
The data are
collected with the Belle detector~\cite{Belle} operating at the
KEKB asymmetric-energy $\EE$ collider~\cite{KEKB}.
This
analysis is based on a 5.7~fb$^{-1}$  $\Upsilon(1S)$ data sample
(102 million $\Upsilon(1S)$ events), a 24.7~fb$^{-1}$
$\Upsilon(2S)$ data sample (158 million $\Upsilon(2S)$ events)~\cite{wangxl},
and a 89.4~fb$^{-1}$ continuum data sample collected at
$\sqrt{s}=10.52$~GeV. Here, $\sqrt{s}$ is the center-of-mass
(C.M.) energy of the colliding $e^+e^-$ system.
The numbers of the $\Upsilon(1S)$ and $\Upsilon(2S)$ events are determined
by counting the hadronic events in the data taken at the $\Upsilon(1S)$ and
$\Upsilon(2S)$ peaks after subtracting the appropriately  scaled
continuum background from the data sample collected at $\sqrt{s} =
9.43$ GeV and 9.993 GeV, respectively. The selection criteria for hadronic events are
validated with the off-resonance data by comparing the measured
$R$ value ($R=\frac{\sigma(\EE\to {\rm hadrons})}{\sigma(\EE\to \MM)}$)
with CLEO's result~\cite{cleoR}.

The {\sc
evtgen}~\cite{evtgen} generator is used to simulate Monte Carlo
(MC) events. For two-body decays, the angular distributions are
generated using the formulae in Ref.~\cite{DPNU}. Inclusive
$\Upsilon(1S)$ and $\Upsilon(2S)$ MC events, produced using {\sc
pythia}~\cite{pythia} with four times the luminosity of the real data,
are used to identify  possible peaking backgrounds from
$\Upsilon(1S)$ and $\Upsilon(2S)$ decays.

The Belle detector is described in detail elsewhere~\cite{Belle}. It is
a large-solid-angle magnetic spectrometer that consists of a
silicon vertex detector (SVD), a 50-layer central drift chamber
(CDC), an array of aerogel threshold Cherenkov counters (ACC),
a barrel-like arrangement of time-of-flight scintillation counters
(TOF), and an electromagnetic calorimeter comprised of CsI(Tl)
crystals (ECL) located inside a superconducting solenoid coil that
provides a 1.5~T magnetic field. An iron flux return located
outside of the coil is instrumented to detect $K_L^0$ mesons and
to identify muons (KLM).


For each charged track other than those from $\ks$ decays, the impact
parameters perpendicular to and along the beam direction with
respect to the interaction point are required to be less than
0.5~cm and 4~cm, respectively, and the transverse momentum must
exceed 0.1~GeV/$c$ in the laboratory frame. Well-measured charged
tracks are selected and the number of good charged tracks must equal
four for the $\ks K^+ \pi^-$ final state or two for the $\pp
\pi^0 \pi^0$ and $\pp\piz$ final states.
For each charged track,
a likelihood ${\mathcal{L}}_X$ is formed from several detector subsystems
for particle hypothesis $X \in \{ e,\ \mu,\ \pi,\ K,\ p \}$.
A track with a likelihood ratio $\mathcal{R}_K = \frac{\mathcal{L}_K} {\mathcal{L}_K
+ \mathcal{L}_\pi}> 0.6$ is identified as a kaon, while a track
with $\mathcal{R}_K<0.4$ is treated as a pion~\cite{pid}. With this
selection, the kaon (pion) identification efficiency is about 85\%
(89\%), while 6\% (9\%) of kaons (pions) are misidentified as
pions (kaons).
Similar likelihood ratios $\mathcal{R}_e$ and $\mathcal{R}_{\mu}$
are defined to identify electrons and muons, respectively~\cite{EID, MUID}.

Except for the $\pp$ pair from $\ks$ decay, all charged tracks
are required to be positively identified as pions or kaons.
The requirements $\mathcal{R}_\mu< 0.95$ and $\mathcal{R}_e<0.95$
for the charged tracks  remove 9.3\% (79\%) of the backgrounds for
$K_S^0 K^+ \pi^-$ ($\pi^+ \pi^- \pi^0$) with no loss in efficiency.

For $\ks$ candidates decaying into $\pp$ in the $\ks K^+ \pi^-$
mode, we require that the invariant mass of the $\pp$
pair lies within a $\pm 8$ MeV/$c^2$
interval around the $\ks$ nominal mass (which contains about 95\%
of the signal according to MC simulation)
and that the pair has a displaced vertex and flight direction
consistent with a $\ks$ originating from the interaction point~\cite{ks}.

To be identified as a photon candidate, a cluster in the electromagnetic calorimeter
should not match the extrapolated position of any charged track and should have energy exceeding
100 (200) MeV in the $\pi^+\pi^-\pi^0\pi^0$ ($\pi^+\pi^-\pi^0$) mode.
A $\pi^0$ candidate is reconstructed from a pair of photons. We perform a
mass-constrained fit to the selected $\pi^0$ candidate and require
$\chi^2<15$.

To remove additional backgrounds in the $\pi^+\pi^-\pi^0$ final state, we require
a matching ECL cluster for each charged track, with
$E_{\pi}^{\rm ECL}/P_{\pi}>0.02$.
Here, $E_{\pi}^{\rm ECL}$
and $P_{\pi}$ represent the energy deposited
in the ECL and the
momentum in the laboratory frame, respectively, for the pion candidate.
To suppress the background events from the initial-state-radiation (ISR) process $\EE
\to \rho^0 \to \pp$ where the charged tracks are combined with a $\pi^0$ candidate,
we require $|(E_1-E_2)/(E_1+E_2)|<0.65$, where $E_1$ and $E_2$ are the
$\pi^0$ daughter-photon energies in the laboratory
frame. To suppress background from the ISR process $\EE \to  \omega \to  \pp \pi^0$,
the same requirement is imposed for the higher-momentum $\pi^0$ in the  $\omega \pi^0$
mode.

\begin{figure}[htbp]
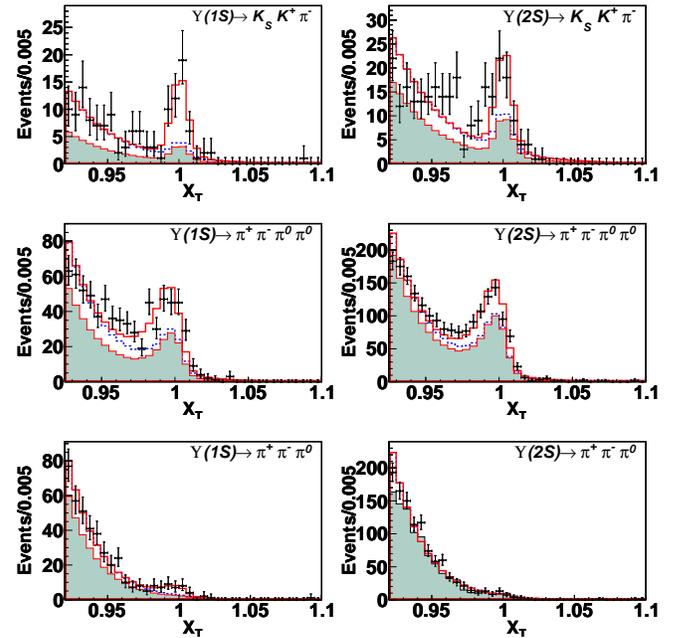

\includegraphics[height=4.2cm,angle=-90]{fig1a.epsi}\vspace{0.25cm}
\includegraphics[height=4.2cm,angle=-90]{fig1b.epsi}\\
\includegraphics[height=4.2cm,angle=-90]{fig1c.epsi}\vspace{0.25cm}
\includegraphics[height=4.2cm,angle=-90]{fig1d.epsi}\\
\includegraphics[height=4.2cm,angle=-90]{fig1e.epsi}
\includegraphics[height=4.2cm,angle=-90]{fig1f.epsi}
\caption{\label{ecms} The fits to the scaled total energy $X_T$
distributions from $\Upsilon(1S)$ and $\Upsilon(2S)$ decays to
$\ks K^+ \pi^-$, $\pp \pi^0 \pi^0$, and $\pp \pi^0$. Solid
points with error bars are from resonance data. The solid histograms
show the best fits, dashed curves are the total background
estimates, and shaded histograms are the normalized continuum
background contributions.}
\end{figure}

We define an energy conservation variable $X_{T} = \Sigma_h
E_h/\sqrt{s}$, where $E_h$ is the energy of the final-state
particle $h$ in the $\EE$ C.M. frame. For signal candidates,
$X_T$ should be around 1. Figure~\ref{ecms} shows the $X_T$
distributions for $\Upsilon(1S)$ and $\Upsilon(2S)$ decays to $\ks
K^+ \pi^-$, $\pp \pi^0 \pi^0$, and $\pp \pi^0$
after applying all selection criteria.
Solid points with
error bars are from data at the indicated $\Upsilon$ resonance.

The continuum background contribution is measured
by extrapolating the data
at $\sqrt{s}=10.52$~GeV to the
$\Upsilon(1S)$ and $\Upsilon(2S)$ resonances. For the
extrapolation, we use the scale factor, $f_{{\rm scale}}=
\frac{\lum_{\Upsilon}}{\lum_{{\rm con}}}
\frac{\sigma_{\Upsilon}}{\sigma_{{\rm con}}}
\frac{\epsilon_{\Upsilon}}{\epsilon_{{\rm con}}}$, where
$\frac{\lum_{\Upsilon}}{\lum_{{\rm con}}}$,
$\frac{\sigma_{\Upsilon}}{\sigma_{{\rm con}}}$ and
$\frac{\epsilon_{\Upsilon}}{\epsilon_{{\rm con}}}$ are the ratios
of luminosity, cross sections and efficiencies, respectively, at the bottomonium
masses and continuum energy points. For nominal results, the $s$
dependence of the cross section is assumed to be
$1/s^3$~\cite{foot} and the corresponding scale factor is
about 0.12 for the $\Upsilon(1S)$ and 0.37 for the $\Upsilon(2S)$.
The dependence of the cross section on the
beam energy could vary from $1/s^3$ to $1/s^4$~\cite{foot, michael};
this range is included as a systematic uncertainty.

Besides the continuum background contribution, we search for possible
backgrounds from $\Upsilon(1S)$ and $\Upsilon(2S)$ decays.
No peaking backgrounds from the $\Upsilon(1S)$ and
$\Upsilon(2S)$ inclusive MC samples are found in the $X_T$ signal
regions. Potential backgrounds due to particle
misidentification --- for example,
from $2(\pi^+ \pi^-)$ and $\kk \pp$ for $\ks K^+ \pi^-$ ---
are estimated and found to be negligible. In the
lower $X_T$ region, backgrounds arise from decays with additional
$\pi^0$'s: from are $\ks K^+ \pi^- \pi^0$ for $\ks K^+ \pi^-$,
$\pp 3\pi^0$ for $\pp \pi^0 \pi^0$, and $\pp \pi^0 \pi^0$ for $\pp
\pi^0$. There are also some backgrounds from $\tau^+ \tau^-\to\pp
n\pi^0 \nu_{\tau} \bar\nu_{\tau}$ with $n\ge 2$ for $\pp \pi^0
\pi^0$, and $n\ge 1$ for $\pp \pi^0$. The $X_T$ distributions from
the above backgrounds are checked with MC simulations and
found to be featureless.

We find that these backgrounds from
$\Upsilon(1S)$ and $\Upsilon(2S)$ decays
together with the normalized contribution from continuum production
can describe the data in
the $X_T<0.975$ region very well.
For $\pi^+ \pi^- \pi^0 \pi^0$
($\pp \pi^0$), the fraction of events with multiple combinations
is 2.1\% (1.7\%) due to multiple $\pi^0$  candidates;
this is consistent with the MC simulation and is taken into
account in the efficiency determination.


An unbinned simultaneous maximum likelihood fit to the $X_T$ distributions
is performed to extract the signal and background yields in the
$\Upsilon(1S)$ and continuum data samples, and in the
$\Upsilon(2S)$ and continuum data samples. The signal shapes are
obtained from MC simulated signal samples directly, where for $\ks K^+ \pi^-$
the signal shape is smeared with a Gaussian function to
account for an 18\% difference in the resolution between data and MC samples.
In this fit, an exponential
background shape is used for the $\Upsilon(1S)$/$\Upsilon(2S)$
decay backgrounds in addition to the normalized continuum
contribution. The fit ranges and results for the $X_T$
distributions from $\ks K^+ \pi^-$, $\pp \pi^0 \pi^0$, and $\pp
\pi^0$ candidate events are shown in Fig.~\ref{ecms}, and the fit
results are summarized in Table~\ref{summary2}.

We determine a Bayesian 90\% confidence level (C.L.) upper limit
on $N_{\rm sig}$ by finding the value $N^{\rm UL}_{\rm sig}$ such
that $ \frac{\int_{0}^{N^{\rm UL}_{\rm sig}} \mathcal{L} dN_{\rm
sig}} {\int_{0}^{\infty} \mathcal{L} dN_{\rm sig}}=0.90, $ where
$N_{\rm sig}$ is the number of signal events and $\mathcal{L}$ is
the value of the likelihood as a function of $N_{\rm sig}$. The
statistical significance of the signal is estimated from the
difference of the logarithmic likelihoods,
$-2\ln(\mathcal{L}_0/\mathcal{L}_{\rm max})$, taking into account
the difference in the number of degrees of freedom in the fits,
where $\mathcal{L}_0$ and $\mathcal{L}_{\rm max}$ are the
likelihoods of the fits without and with signal, respectively.

\begin{figure}[htbp]
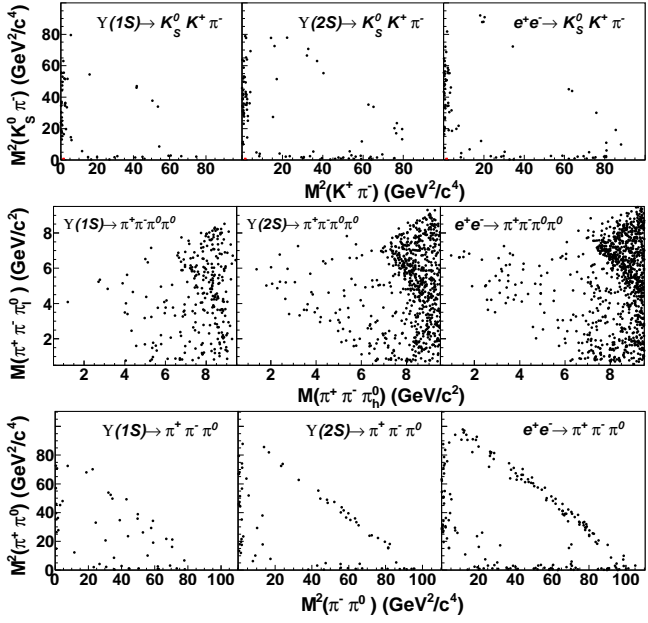

\includegraphics[height=8.5cm,angle=-90]{fig2a.epsi}\\
\includegraphics[height=8.5cm,angle=-90]{fig2b.epsi}\\
\includegraphics[height=8.5cm,angle=-90]{fig2c.epsi}
\caption{\label{dalitz} Dalitz plots for the $\ks K^+ \pi^-$ (top row)
and $\pp \pi^0$ (bottom row) final states, and scatter plot for the
$\pp \pi^0 \pi^0$ (middle row) final state. Here, the left column
is for $\Upsilon(1S)$ decays, the middle column is for
$\Upsilon(2S)$ decays, and the right column is for the continuum
data without normalization. In $\pp \pi^0 \pi^0$, $\pi^0_h$ and $\pi^0_l$ represent the
pion with a higher and lower momentum in the laboratory system,
respectively.}
\end{figure}

After requiring the value of the variable  $|X_T-1|$ to be less than
0.02 for $\ks K^+ \pi^-$
and less than 0.025 for $\pp \pi^0 \pi^0$ and $\pp\piz$, the
Dalitz plots for the $\ks K^+ \pi^-$ and $\pp \pi^0$ final states and
the scatter plot of $M(\pp \pi^0_l)$ versus $M(\pp\pi^0_h)$ for
the $\pp \pi^0_h \pi^0_l$ final state are shown in
Fig.~\ref{dalitz}. In the scatter plot, $\pi^0_h$ and $\pi^0_l$
represent the pion with a higher and lower momentum in the
laboratory system, respectively. According to MC simulated
$\Upsilon\to \omega \pi^0$ signal events, over 97\% of the $\pi^0$s
from $\omega$ decays have the lower momentum and there is
only one $\pp\pi^0$ combination in the $\omega$ mass region.

\begin{figure}[htbp]
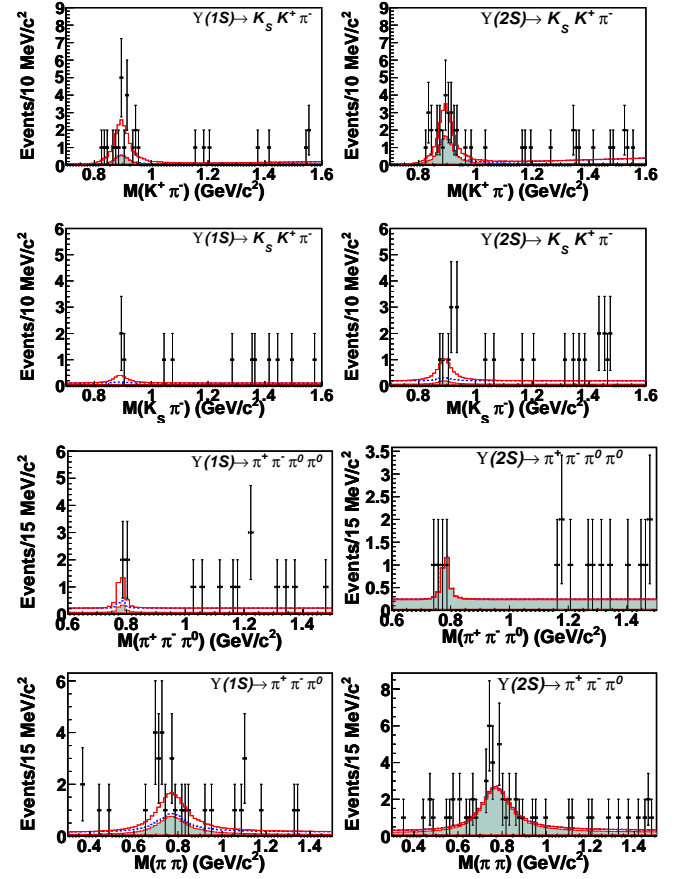

\includegraphics[height=4.2cm,angle=-90]{fig3a.epsi}\vspace{0.25cm}
\includegraphics[height=4.2cm,angle=-90]{fig3b.epsi}\\
\includegraphics[height=4.2cm,angle=-90]{fig3c.epsi}\vspace{0.25cm}
\includegraphics[height=4.2cm,angle=-90]{fig3d.epsi}\\
\includegraphics[height=4.2cm,angle=-90]{fig3e.epsi}\vspace{0.25cm}
\includegraphics[height=4.2cm,angle=-90]{fig3f.epsi}\\
\includegraphics[height=4.2cm,angle=-90]{fig3g.epsi}
\includegraphics[height=4.2cm,angle=-90]{fig3h.epsi}
\caption{\label{vt-fit} The fits to the $K^+ \pi^-$, $\ks \pi^-$,
$\pp \pi^0$ and $\pi \pi$ mass distributions for the $K^{\ast
}(892)^0$, $K^{\ast}(892)^-$, $\omega$ and $\rho$ vector meson
candidates from $\ks K^+ \pi^-$, $\pp \pi^0 \pi^0$ and $\pp \pi^0$
events from $\Upsilon(1S)$ and $\Upsilon(2S)$ decays (VP modes).
The solid histograms show the results of the simultaneous fits, the
dotted curves show the total background estimates, and the shaded
histograms are the normalized continuum contributions.}
\end{figure}

For the selected events, Fig.~\ref{vt-fit} shows the $K^+ \pi^-$
and $\ks \pi^-$ invariant mass distributions for the $\ks K^+ \pi^-$
final state, the $\pp \pi^0$ invariant mass distribution for
the $\pp\piz\piz$ final state, and
the $\pi\pi$ invariant mass
distribution for the $\pp\piz$ final state~\cite{mpipi}. There are hints of the
vector mesons $K^{\ast }(892)^0$, $K^{\ast }(892)^-$, $\omega$, and
$\rho$ in the expected mass regions,
but except for possible evidence for a $K^{\ast }(892)^0$ signal from $\Upsilon(1S)$  decays,
there is no indication of signal in any other final state.

We perform a similar unbinned simultaneous maximum likelihood fit described
above for  $X_T$ distributions, except that a
first-order Chebyshev polynomial is used instead of the exponential background shape.
Because of the limited statistics, in the fits we assume there is no interference between
the vector meson signal and other non-resonant components. We also neglect
the possible small interference between the $\Upsilon$ resonance decays
and continuum process due to the narrow widths
of the $\Upsilon$ resonances.
The results of the fits  are shown in
Fig.~\ref{vt-fit} and listed in Table~\ref{summary2}.

\begin{sidewaystable}
\caption{Results for the $\Upsilon(1S)$ and $\Upsilon(2S)$ decays,
where $N_{\rm sig}$ is the number of signal events from the fits, $N^{\rm
UL}_{\rm sig}$ is the upper limit on the number of signal events,
$\epsilon$ is the efficiency (\%), $\Sigma$ is the statistical
significance ($\sigma$), $\BR$ is the branching fraction
(in units of $10^{-6}$),
$\BR^{\rm UL}$ is the 90\% C.L. upper limit on the branching fraction,
$Q_{\Upsilon}$ is the ratio of the $\Upsilon(2S)$ and
$\Upsilon(1S)$ branching fractions, and $Q_{\Upsilon}^{\rm UL}$ is
the upper limit on the value of $Q_{\Upsilon}$. The first error in $\BR$ and $Q_{\Upsilon}$ is
statistical, and the second systematic. Here $\BR(\ks \to \pp)$
has been included in the efficiency for the $\ks K^+ \pi^-$ final
states. In order to set conservative upper limits on these
branching fractions, the efficiencies are lowered by a factor of
$1-\sigma_{\rm sys}$ in the calculation, where $\sigma_{\rm sys}$
is the total systematic error.} \label{summary2}
\begin{center}{
\begin{tabular}{c|llcccl|llcccl|cc}
\hline
Channel & \multicolumn{6}{c|}{$\Upsilon$(1S)}
         &  \multicolumn{6}{c|}{$\Upsilon$(2S)}   &   \\
        & $N_{\rm sig}$ & $N^{\rm UL}_{\rm sig}$& $\epsilon$ & $\Sigma$ & $\BR$
        &$\BR^{\rm UL}$  & $N^{\rm sig}$ & $N^{\rm UL}_{\rm sig}$ & $\epsilon$
        & $\Sigma$ &$\BR$ &$\BR^{\rm UL}$  &$Q_{\Upsilon}$ & $Q_{\Upsilon}^{\rm UL}$
\\
\hline \rule{0mm}{0.4cm}
 $\ks K^+ \pi^-$ & $37.2\pm 7.6$&---&22.96&6.2&$1.59\pm0.33\pm0.18$&---
                 & $39.5\pm 10.3$&---&21.88&4.0&$1.14\pm0.30\pm 0.13$ & ---
                 & $0.72\pm0.24\pm0.09$  & --- \\
 $\pp \pi^0 \pi^0$&$143.2\pm22.4$&---& 11.20&7.1&$12.8\pm2.01\pm2.27$&---
                  & $260.7\pm37.2$&---&12.98&7.4&$13.0\pm1.86\pm2.08$&---
                  &$1.01\pm0.22\pm0.23$  & --- \\
 $\pp \pi^0$ & $25.5\pm 8.6$ & ---  & 11.86 & 3.4 & $2.14\pm0.72\pm 0.34$  & ---
             & $-2.1 \pm 9.5$ & 15 & 13.19 & ---& $-0.10\pm0.46 \pm 0.02$ & 0.80
             & $-0.05\pm0.21\pm0.02$   & 0.42 \\\hline
 $K^{\ast}(892)^0\bar{K}^0$&$16.1\pm4.7$&---&16.23& 4.4&$2.92\pm0.85\pm0.37$&---
           &$14.7\pm6.0$&30&15.59&2.7& $1.79\pm0.73\pm0.30$ &4.22
           &$0.61\pm0.31\pm0.12$&1.20\\
 $K^{\ast}(892)^- K^+$&$2.0\pm1.9$& 6.3& 18.92&1.3&$0.31\pm0.30\pm0.04$&1.11
           &$5.7\pm3.4$&13&18.77&2.0&$0.58\pm0.35\pm0.09$&1.45
           &$1.87\pm2.12\pm0.33$&5.52\\
 $\omega \pi^0$&$2.5\pm2.1$& 6.8& 2.11& 1.6&$1.32\pm1.11\pm0.14$
               &3.90&$0.1\pm2.2$&4.6&2.32&0.1&$0.03\pm0.68\pm0.01$&1.63
               &$0.02\pm0.50\pm0.01$& 1.68  \\
 $\rho \pi$  & $11.3\pm5.9$  & 22  & 6.41 & 2.2 & $1.75\pm0.91\pm 0.28$  & 3.68
        & $-1.4\pm8.6$  & 14  & 8.66 & --- & $-0.11\pm0.64\pm 0.03$ &1.16
        & $-0.06\pm 0.38 \pm 0.02$  & 0.94  \\
\hline
\end{tabular}
}
\end{center}
\end{sidewaystable}


There are several sources of systematic errors for the branching
fraction measurements. The uncertainty in the tracking efficiency
for tracks with angles and momenta characteristic of signal events
is about 0.35\% per track and is additive. The uncertainty due to
particle identification efficiency is 1.7\% with an efficiency
correction factor of 0.98 for each pion, and is 1.6\% with an
efficiency correction factor of 0.97 for each kaon. The uncertainty
in selecting a $\pi^0$ candidate is estimated
using a control sample of $\tau^- \to
\pi^- \pi^0 \nu_{\tau}$ events.
We include a 2.2\% systematic
error with efficiency correction factors of 0.94 for low-momentum
and 0.97 for high-momentum $\pi^0$ mesons.
In the $\ks K^+ \pi^-$
mode, the $K_S^0$ reconstruction and the systematic error is verified by comparing the
ratio of $D^+\to K_S^0\pi^+$ and $D^+\to K^-\pi^+\pi^+$ yields
with the MC expectations;
the difference between data and MC
simulation is less than 4.9\%~\cite{ks-error}.
The efficiency of the requirement $E_{\pi}^{\rm
ECL}/P_{\pi}>0.02$  is 97.4\% in $\pp \pi^0$ and
the uncertainty can be neglected according to a check of the results
with and without this requirement.
Errors on the
branching fractions of the intermediate states are taken from the
PDG listings~\cite{PDG}.
For the $\pi^+ \pi^- \pi^0$ final state, the trigger efficiency is verified
using the pure ISR control sample $e^+e^- \to \omega \to \pi^+ \pi^- \pi^0$.
According to MC simulation,
for the $\pi^+ \pi^- \pi^0$ ($\rho \pi$) mode, the trigger efficiency is
97\% (94\%), with an uncertainty that is smaller than 1.5\% (3\%);
for the other modes, the
trigger efficiency is greater than 99\% and the corresponding
uncertainty is neglected. The trigger efficiency in $\rho \pi$
is somewhat lower due to high momentum $\pi^0$ in $\rho^0 \pi^0$.
We estimate the systematic errors
associated with the fitting procedure by changing the order of the
background polynomial, the range of the fit and introducing an extra
Gaussian function to describe the possible excess around $X_T\sim 0.96$ in $X_T$ fits
and take the differences in
the results of the fits, which are 1.5\%-11\% depending on the final
state particles, as systematic errors.
To investigate the effect of possible intermediate resonances for the
$\Upsilon(1S)$/$\Upsilon(2S)$ $\to \ks K^+ \pi^-$, $\pp \pi^0 \pi^0$ and $\pp \pi^0$ decays,
the efficiencies are estimated
by using sampled phase space MC signal events according to the Dalitz plot or scatter plot
that are shown in Fig.~\ref{dalitz}.
The difference is 7.3\%/5.3\% for $\Upsilon(1S)/\Upsilon(2S)
\to \ks K^+ \pi^-$, 5.6/4.4\% for $\Upsilon(1S)/\Upsilon(2S)
\to \pp \pi^0 \pi^0$, 11\%/7.8\% for $\Upsilon(1S)/\Upsilon(2S)
\to\pp \pi^0$; these values are assigned as a systematic
uncertainty due to this source.
For the $K^{\ast}(892) {\bar K}$ and $\rho \pi$ modes, we
estimate the systematic errors associated with the resonance
parameters by changing the values of the masses and widths of the
resonances by $\pm 1\sigma$. The $K^{\ast}(892)$ and $\rho$
line shapes are replaced by a
relativistic Breit-Wigner function and the Gounaris-Sakurai
parametrization~\cite{GS}, respectively. The total differences of
2.6\%-11\% in the fitted results are taken as systematic errors.
For the central values of the branching fractions, the
difference between alternative C.M. energy dependences of the
cross section is included as a systematic error due to the
uncertainty of the continuum contribution, which is in the range
of 4.7\% to 22\%. The uncertainty due to limited MC statistics is
at most 2.7\%. Finally, the uncertainties on the total numbers of
$\Upsilon(1S)$ and $\Upsilon(2S)$ events are 2.2\% and 2.3\%,
respectively, which are mainly due to imperfect simulations
of the charged multiplicity distributions from inclusive hadronic
MC events. Assuming that all of these systematic error sources
are independent, the total systematic error is 11\%-26\% depending
on the final state, as shown in Table~\ref{totalsys}.

\begin{table*}[htbp]
\caption{Relative systematic errors (\%) on the decay
branching fractions. } \label{totalsys}{\footnotesize
\begin{tabular}{l | c c c | c c c c }
\hline Source ($\Upsilon(1S)/\Upsilon(2S)$) & $\ks K^+ \pi^-$  & $\pp \pi^0 \pi^0$ &  $\pp \pi^0$ & $K^{\ast}(892)^0 \bar{K}^0$  & $K^{\ast}(892)^- K^+$& $\omega \pi^0$ & $\rho \pi$ \\\hline
 Tracking & 0.7 & 0.7 & 0.7 & 0.7 & 0.7 & 0.7 & 0.7 \\
 PID      & 3.3 & 3.4 & 3.4 & 3.3 & 3.3 & 3.4 & 3.4  \\
 $\pi^0$ selection & --- & 4.4 & 2.2 & --- & --- & 4.4 & 2.2 \\
 $\ks$ selection & 4.9 & --- & --- & 4.9 & 4.9 & --- & ---  \\
 Branching fractions & 0.1 & 0.1 & 0.1 & 0.1 & 0.1 & 0.8 & 0.1 \\
 Trigger & --- & --- &  1.5  & --- & --- & --- & 3.0 \\
 Fitting procedure & 1.5/3.4 & 3.7/5.6 & 9.3/9.5 &  8.0/11 & 9.4/11 & 4.9/11 & 3.2/5.3  \\
 Intermediate resonance & 7.3/5.3 & 5.6/4.4 &  11/7.8  & --- & --- & --- & --- \\
 Resonance parametrization & --- &--- & --- &  2.6/3.8 & 4.0/6.2 & --- & 4.6/11 \\
  Continuum uncertainty  & 4.7/4.7 & 15/12 & 4.7/12 &  6.5/9.4 & 3.7/1.4 & 6.8/6.6  & 14/22 \\
 MC statistics & 1.7/1.8 &2.7/2.5 & 0.9/0.8 &  2.2/2.3 & 2.1/2.1 & 2.0/1.9 & 0.9/0.8 \\
 Number of $\Upsilon$ events & 2.2/2.3 & 2.2/2.3 & 2.2/2.3 & 2.2/2.3 & 2.2/2.3 & 2.2/2.3 & 2.2/2.3 \\
 \hline
 Sum in quadrature & 11/11 & 18/16 & 16/18 & 13/17& 13/15 & 11/15 &  16/26 \\\hline
\end{tabular}}
\end{table*}


Table~\ref{summary2} shows the  results for the branching
fractions including the upper limits at 90\% C.L. for the channels
with a statistical significance of less than 3$\sigma$. In order to set conservative upper
limits on these branching fractions, the efficiencies are lowered
by a factor of $1-\sigma_{\rm sys}$ in the calculation, where
$\sigma_{\rm sys}$ is the total systematic error.
The corresponding ratio of the branching fractions  of $\Upsilon(2S)$
and $\Upsilon(1S)$ decay ($Q_{\Upsilon}$) is calculated; in some cases, the
systematic errors cancel. A Bayesian upper limit
on the ratio at the 90\% C.L. ($Q_{\Upsilon}^{UL}$)
is obtained by performing toy MC experiments. 
We sample $\BR_{\Upsilon(1S)}$ and $\BR_{\Upsilon(2S)}$
by assuming they follow Gaussian distributions, where the mean values and standard 
deviations of the Gaussian functions
are set to be the central value and total error (with a common error removed) of the branching fraction, respectively.
For the sampled distribution of the ratio of the branching fractions greater than zero, we obtain $Q_{\Upsilon}^{UL}$,
where $Q_{\Upsilon}^{UL}$ corresponds to the number of experiments with 
$Q_{\Upsilon}<Q_{\Upsilon}^{UL}$ in less than 90\% of the total number of toy experiments.
 At present, all the results
on the branching fractions, including upper limits reported in this letter, are the first
measurements or the best measurements.

In summary, we have measured $\Upsilon(1S)$ and $\Upsilon(2S)$
exclusive hadronic decays to $\ks K^+ \pi^-$, $\pi^+ \pi^- \pi^0
\pi^0$, and $\pi^+ \pi^- \pi^0$, as well as the two-body VP
($K^{\ast}(892)^0\bar{K}^0$, $K^{\ast}(892)^-K^+$, $\omega\pi^0$, and
$\rho\pi$) states. Signals are observed for the first time in the
$\Upsilon(1S)\to \ks K^+ \pi^- $, $\pp \pi^0 \pi^0$ and
$\Upsilon(2S) \to \pp \pi^0 \pi^0$ decay modes.
Although many $\Upsilon(1S)$ and $\Upsilon(2S)$ exclusive decay modes
were previously measured using CLEO data~\cite{seth}, only the
$\ks K^+ \pi^-$ mode overlaps with our measurement and upper limits at 90\%  C.L.
were presented there. Our results for the $\ks K^+ \pi^-$ mode are
well below the upper bounds
reported in Ref.~\cite{seth}.
There is an indication for large
isospin-violation between the branching fractions for the charged
and neutral $ K^{\ast}(892) {\bar K}$ for both $\Upsilon(1S)$ and
$\Upsilon(2S)$ decays, as in $\psp$ decays, which indicates
that the electromagnetic process plays an important role in these decays~\cite{isospin}.
We find that, for the processes $\ks K^+ \pi^- $ and $\pi^+ \pi^- \pi^0 \pi^0$,
the $Q_{\Upsilon}$ ratios are consistent with the expected value;
for $\pp \pi^0$, the $Q_{\Upsilon}$ ratio is a little lower
than the pQCD prediction. The results for the other modes are
inconclusive due to low statistical significance. These results
may supply useful guidance for interpreting violations of the 12\%
rule for OZI-suppressed decays in the charmonium sector.


We thank the KEKB group for excellent operation of the
accelerator; the KEK cryogenics group for efficient solenoid
operations; and the KEK computer group, the NII, and
PNNL/EMSL for valuable computing and SINET4 network support.
We acknowledge support from MEXT, JSPS and Nagoya's TLPRC (Japan);
ARC and DIISR (Australia); FWF (Austria); NSFC (China); MSMT (Czechia);
CZF, DFG, and VS (Germany);
DST (India); INFN (Italy); MEST, NRF, GSDC of KISTI, and WCU (Korea);
MNiSW and NCN (Poland); MES and RFAAE (Russia); ARRS (Slovenia);
IKERBASQUE and UPV/EHU (Spain);
SNSF (Switzerland); NSC and MOE (Taiwan); and DOE and NSF (USA).



\begin{thebibliography}{**}
\bibitem{charge} Charge-conjugate decays are implicitly assumed
throughout the paper and the $\rho \pi$ mode includes $\rho^0 \pi^0$,
$\rho^+ \pi^-$ and $\rho^- \pi^+$.
\bibitem{isospin} The ratio of cross sections between neutral and charged
$ K^{\ast}(892) {\bar K}$ is larger than 4.6 at 90\% C.L. in continuum data sample
at $\sqrt{s}=10.52$ GeV, which is not very different from the expected value of 4
with the assumption of $SU(3)$ symmetry~\cite{su3}.
\end{thebibliography}
\end{document}